\documentstyle[12pt]{article}

\begin{document}
\baselineskip=18.6pt plus 0.2pt minus 0.1pt \makeatletter
\@addtoreset{equation}{section} \renewcommand{\theequation}{\thesection.%
\arabic{equation}} \makeatletter \@addtoreset{equation}{section}
\date{}
\begin{titlepage}

\title{\vspace{-3cm}
\hfill\parbox{4cm}{}\\
 \vspace{1cm}
 Insight on confinement using scalar field interactions}
 \vspace{2cm}

\author{ R. Markazi$^{1,2}$\footnote{r.markazi@uiz.ac.ma}  and   N. El Biaze$^{1,2}$\footnote{n.elbiaze@uiz.ac.ma}\\
\
\small{$^1$ Materials and renewable energies laboratory, Faculty of science, Ibn Zohr University-Agadir, Morocco.}
\\ \small{$^2$ High School of Technology-Guelmim, Ibn Zohr University-Agadir, Morocco.}} \maketitle \setcounter{page}{1}

\begin{abstract}
The scalar field plays an fundamental role in the investigation of
confinement property characterising many particle physics models. This is achieved
by coupling this particle  directly with gauge fields at the lagrangian level.
We have adopted the same approach to determine a potential {[}10{]}
as a perturbative series in terms of interquark distance. In order
to introduce the gravitational effects and inspired from bag models,
we implement a scalar field which interacts both with the vacuum and
the electron field. In this context and with presence of the vacuum
condensates, it is possible to derive a more accurate expression of
the electron energy.

\end{abstract}
\end{titlepage}

\newpage
\section{Introduction}

Bag model is an interesting approach which provides an explanation
of confinement property characterising quarks interactions. They are
also used to describe some physical phenomena related to the stability
of some particles like electron. According to this approach, one supposes
the existence of a cavity, which is a space zone filled with scalar
condensate.The scalar field appears in many theoretical physics models
with different meanings. Sometimes it appears as a massless field
and in others as a massive one.
In QCD models, it is seen as a pseudo-Goldstone boson{[}5{]}. Whereas
in cosmology, it is an hypothetical particle considered as a dark
matter candidate that constitutes the missing mass of universe {[}6{]}.
The scalar field can couples in different ways. For example in string
theory, it couples to super-Yang Mills gauge fields in curved space
whereas in the Brans-Dicke model, it couples to the Ricci curvature{[}8{]}.
For a long time the electron stability was studied using different
approaches. These latter faced several difficulties related to the
electron finite size, the origin of its mass and the problem with
the relativistic transformation properties of the energy and momentum
of the electromagnetic field of the electron. To overcome these difficulties,
one introduces a model of an electron based on a charged conducting
surface of a cavity in the electromagnetic field. In this way, we
get phenomenological solution to the electron stability problem. In
fact, the cavity exhibits a surface tension due to the difference
of the condensate density between its inside and its outside.

The surface tension depends on the parameters defining the Higgs potential
in the electroweak gauge-model theory {[}9-14{]}. It allows to establish
the expression of the total energy of the electron and its surrounding
cavity. The total energy depends on the vacuum expectation value of
the Higgs field and the Ginzburg-Landau coherence length {[}15{]}.

\section{Confining potential from scalar field-gluon coupling}

The confinement property means that the quarks and gluons cannot exist
as separate objects. To investigate it, various QCD models were proposed
to derive an interquark potential which exhibits confinement behavior.
Such potential can be obtained by considering the coupling of a dilaton
$\phi$ to the $4d SU(Nc)$ gauge field like in string theory models{[}8{]}.
To this end, one suggests the following effective lagrangian:
\begin{equation}
L=-\frac{1}{4F(\phi)}G_{\mu\nu}G^{\mu\nu}-\frac{1}{2}\partial_{\mu}\phi\partial^{\mu}\phi-\frac{1}{2}m^{2}\phi^{2}+J_{\mu}^{a}A_{a}^{\mu},
\end{equation}
where the coupling $\frac{1}{4F(\phi)}$ is a function
of the dilaton field $\phi$ and $m$ is the dilatonic mass. The form
of this coupling can have several expressions according to the theoretical
frameworks. The current density $J_{a}^{\nu}=g\delta(r)C_{a}\eta^{\mu}$ appearing
in the lagrangian is considered as a point like static source. $C_{a}$
is the expectation value of the $SU(N_{c})$ generators for a normalized
spinor in the color space, satisfying the algebra identity $\Sigma_{a}C_{a}^{2}=\frac{N_{c}^{2}-1}{2N_{c}}$.\\
Using the lagrangian {(2.1)} , one derives the following equations
of motion corresponding to $\phi$ and $A_{\mu}$ fields:
\begin{equation}
\partial^{2}\phi-m^{2}\phi=\frac{1}{4}\frac{d(1/F(\phi))}{d\phi}G_{\mu\nu}G^{\mu\nu}
\end{equation}

\begin{equation}
 \partial_{\mu}(\frac{1}{F(\phi)}G_{a}^{\mu\nu})+g\frac{1}{F(\phi)}A_{\mu}^{b}f_{ab}^{c}G_{c}^{\mu\nu}=-J_{a}^{\nu},
\end{equation}

which can be rewritten as:
\begin{equation}
\frac{d^{2}}{dr^{2}}(r\phi)-m^{2}(r\phi)=\frac{\mu2}{2r^{3}}\frac{d(F(\phi))}{d\phi}
\end{equation}
where $\mu=\frac{g}{4\pi}\sqrt{\frac{Nc^{2}-1}{2N_{c}}}$.\\
The equation (2.4) may be solved for a given dilaton-gluon coupling
$F(\phi)$ for which the interquark potential $V(r)$ is given by
the following formula {[}10{]}:
\begin{equation}
 V(r)=-\frac{g}{4\pi}C\intop\frac{F(\phi(r))}{r^{2}}dr.
\end{equation}

This expression of the potential is very interesting as it generalizes
the Coulombian potential formula $V_{c}(r)\sim\frac{1}{r}$, recovered
in the particular case: $F(\phi)=1$. The main reason for introducing
the coupling between the dilaton field and the Yang Mills field strength
is to have an alternative way to derive more general potential which
takes into account the confinement behavior of interquark interaction.
Following this approach, one can get many forms of the interquark potential
just by modifying the dilaton-gluon coupling $F(\phi)$. We have shown
in {[}8,12{]} that the confining terms are
directly related to the QCD vacuum condensates describing the nonperturbative
effects.

\section{Influence of scalar field condensate on electron energy }

{In order to investigate the electron stability, a model
{[}10{]}based on a scalar field $\phi$ and a $U(1)$ gauge field
$A_{\mu}$ was suggested. The electron charge is supposed to be uniformly
distributed on of a spherical cavity immerged in a vacuum filled with
the Higgs condensate. In this approach, the Lagrangian can be written
as:
\begin{equation}
{\mathcal{L}}={\mathcal{L}}^{\phi}+{\mathcal{L}}^{E}
\end{equation}
Where the scalar part of the lagrangian is:

\begin{equation}
{\mathcal{L}^{\phi}}=\frac{1}{2}g_{\mu\nu}\partial \phi_{\mu} \partial^{\nu}\phi+V(\phi)
\end{equation}

the conventional Minkowski metric tensor
$g_{\mu\nu}=(1,-1,-1,-1)$.}
{To ensure the spontaneous symmetry breaking, $V(\phi)$
is chosen as {[}11{]}: {$V(\varphi)=-2\mu^{2}\phi^{2}+\frac{1}{4}\lambda\phi^{4}$}.\\
{The electric charge of the electron gives rise to
the Yang-Mills term :
\begin{equation}
{\mathcal{L}}^{E}=-\frac{1}{16\pi}F_{\alpha\beta}F^{\alpha\beta}
\end{equation}
where $F_{\alpha\beta}=\partial_{\alpha}A_{\beta}-\partial_{\beta}A_{\alpha}$
is the field-strength tensor.}
In this framework, the stress energy tensor given by the the relation:
\begin{equation}
T_{ij}=\frac{2}{\sqrt{\left|g\right|}}\frac{ \partial}{\partial g^{ij}}(\sqrt{\left|g\right|}\mathcal{L})
\end{equation}
can be splitted into two main parts:
\begin{equation}
T_{ij}=T^{H}_{ij}+T^{E}_{ij}
\end{equation}
where $T_{ij}^{H}$ and $T_{ij}^{E}$ are respectively the scalar and
the gauge field contributions:
\begin{equation}
T_{ij}^{H}=\partial_{i}\phi\partial_{j}\phi-g_{ij}\mathcal{L}^{H}
\end{equation}
and
\begin{equation}
T_{ij}^{E}=\frac{1}{4}g_{\beta j}F_{i\lambda}F^{\lambda\beta}+\frac{1}{4}g_{ij}F_{\mu\lambda}F^{\mu\lambda}
\end{equation}
Let's consider the electron charge as a static source confined in
a spherical cavity of radius $R$ and that the scalar field vanishes
inside it whereas, it is supposed to have a fixed value at infinity.
A coherent length denoted by $\delta$ was defined in order to determine
the transitional space regions which separates between the two different
vaccum expectation values of the scalar field. This transitional zone
can be seen as a domain wall separating the false vaccum and the true
vaccum regions. With these assumptions, the total energy can be written as follows$[10]$.
\begin{equation}
E_{tot}(R)=\frac{e^{2}}{2R}+4\pi\sigma R^{2}
\end{equation}
where $\sigma$ is the surface tension due to the scalar field expectation
value outside the cavity $\eta$ given by: $\sigma=\frac{2\eta^{2}}{3\delta}$. \\
The first term appearing in (3.8) represents the coulomb energy
of the cavity surface whereas the second term arises from both the
presence of the condensation energy and the coherent length.

\section{Gravitational effects on electron energy}

The two approaches mentioned above provides two different ways to
derive an interaction potential which is able to take into account
both the colombian interaction and the confinement behavior.

In the first approach, a direct coupling between the scalar field
and the field strength is necessary to obtain such interaction potential
whatever is the form of the scalar field space time distribution.

In the second approach, more interest was given to the scalar spacetime
distribution with the absence of interaction between the scalar field
and the electric field.

Till now, we did not take in consideration the contribution of the
gravitational effects to the total energy of the electron. Motivated
by their importance, we extend the two above approaches by adding
a new term to the lagrangian which describes these effects as an interaction
between the scalar field and the space-time curvature. In this framework,
the scalar field lagrangian becomes:

\begin{equation}
{\mathcal{L}}^{\phi}=-\frac{1}{2}(\mu^{2}+\xi R)\phi^{2}-\frac{\lambda}{4}\phi^{4}
\end{equation}

Where the coupling between the scalar field
$\phi$ and the space-time curvature $R$ is measured by the $\xi$
parameter. We retrieve the previous form of the potential in the
particular case {[}11{]} $\xi=0$. where $\mu$
and $\lambda$ are standard parameters.

In our case, the scalar field vaccum expectation value obtained through
the minimisation of the potential (4.1), is $\eta^{2}=\frac{\mu^{2}+\xi R}{\lambda}$.\\
Using this latter, the potential can be rewritten like:

\begin{equation}
V(\phi)=\frac{\lambda}{4}(\phi^{2}-\eta^{2})^{2}-\frac{\lambda}{4}\eta^{4}
\end{equation}

Let's suppose that the cavity radius is an approximation of the space-time curvature. Consequently, the total energy of the electron becomes:

\begin{equation}
E_{tot}=\frac{e^{2}}{2R}+\frac{\mu^{2}}{\delta\lambda}\pi R^{2}+\frac{\xi}{\delta\lambda}R^{3}
\end{equation}

By analyzing (4.3), we remark that the curvature nature of space
time brings a new contribution to the confining part of the total
electron energy.

Finally, due to the space-time curvature coming from the gravitational
effects, the value of the condensation energy which garantees the
confinement behaviour of our model is $R$ dependent:
\begin{equation}
\epsilon_{cond}=-\frac{1}{4\lambda^{3}}\left(\mu^{2}+\xi R\right)^{2}
\end{equation}
.

Because particles are not free but instead live in a small space time
region, any tentative to get an estimation, for example of the electron
radius must take into account both the gravitational and the vaccum
effects.

\section*{Acknowledgements}
We would like to thank A. Ihlal and A. Rachidi for their support.


\begin{thebibliography}{99}\footnotesize

\bibitem{1} A. Chodos, R. J. Jaffe, K. Johnson, C. B. Thorm and V. F.
Weisskopf, Phys. Rev. D9, 3471 (1974).

\bibitem{2} W. A. Bardeen, M. S. Chanowitz, S. D. Drell, M. Weinstein
and T. M. Yan, Phys. Rev. D11, 1094 (1975).

\bibitem{3} P. A. M. Dirac, Proc. Roy. Soc. 268A, 57 (1962).

\bibitem{4} N. Seiberg and E. Witten, Nucl. Phys. B426 (1994) 19.

\bibitem{5} A. Sakharov, JETP Lett. 5 (1967) 24.

\bibitem{6} R. Dick, Euro. Phys. J. C6 (1999) 701. R. Dick and L.P. Fulcher,
Euro. Phys. J. C9 (1999) 271.

\bibitem{7} S. Weinberg, Phys. Rev. Lett. 19, 1264 (1967). {[}10{]} A.
Salam, in Elementary Particle Theory, ed. W. Svartholm, (Almquist
and Wiskell, Stockholm (1968)).

\bibitem{8} P. D. B. Collins, A. D. Martin and E. J. Squires, Particle
Physics and Cosmology, (John Wiley \& Sons, New York (1989)).

\bibitem{9} A. L. Fetter and J. D. Walecka, Quantum Theory of Many-particle
Systems, (McGraw-Hill, New York (1971)).

\bibitem{10} E. Simanek "Stability of an electron embedded in Higgs
condensate" 	arXiv:1502.00983
\bibitem{11}E. Bentivegna, V. Branchina, F. Contino, D. ZappalàImpact of New Physics on the EW vacuum stability in a curved spacetime background arXiv:1708.01138
\bibitem{12} M. Chabab, R. Markazi and E. H. Saidi, Euro. Phys. J. C13
(2000) 543;
\bibitem{13} M. Chabab, N.EL Biaze, R. Markazi and E. H. Saidi, Class.Quant.Grav.18:5085-5096,2001.
\bibitem{14} C. Barcelo, S. Liberati and M. Visser, Analog gravity from eld theory normal modes,
arXiv:gr-qc/0104001.

\bibitem{15} J. D. Jackson, Classical Electrodynamics, (John Wiley \&
Sons, Inc., New York (1975)).

\bibitem{16} T. D. Lee, Particle Physics and Introduction to Field Theory,
(Harwood Academic Publishers, Chur, Switzerland (1981)).

\end{thebibliography}
\end{document}